\documentclass[preprint,aps,showpacs]{revtex4}

\newcommand{\sigmap}{{\sigma^\prime}}
\newcommand{\sigmaz}{{\sigma_0}}
\newcommand{\ssp}{{\sigma\to\sigmap}}
\newcommand{\sps}{{\sigmap\to\sigma}}

\usepackage{graphicx}

\begin{document}

\title{A Definition of Metastability for Markov Processes with Detailed Balance}

\author{
Fran\c{c}ois Leyvraz, Hern\'{a}n Larralde and David P.~Sanders
}

\affiliation{
Centro de Ciencias F\'\i sicas, UNAM; Apartado postal 48-3, 
CP 62251 Cuernavaca Morelos, Mexico 
}
\begin{abstract}
 A definition of metastable states applicable to arbitrary finite
 state Markov processes satisfying detailed balance is discussed. In
 particular, we identify a crucial condition that distinguishes
 genuine metastable states from other types of slowly decaying modes
 and which leads to properties similar to those postulated in the
 restricted ensemble approach \cite{pen71}.  The intuitive physical
 meaning of this condition is simply that the total equilibrium
 probability of finding the system in the metastable state is
 negligible. As a concrete application of our formalism we present
 preliminary results on a 2D kinetic Ising model.
\end{abstract}
\maketitle 

\section{Introduction} 

In equilibrium statistical mechanics, it has been accepted for a long
time that the canonical ensemble provides, at least for the vast
majority of systems, an adequate description of their properties. It
could be argued that this essentially reduces the problem of
understanding equilibrium properties to one of computation. On the
other hand, for statistical mechanics far from equilibrium things are
quite different: there are no laws corresponding to the Gibbs ensembles
in order to calculate the probabilities of a given configuration in
general, and the situation is still extremely complex.

In this respect, metastable states occupy a curiously intermediate
position: they are generally viewed as equilibrium states, i.e.\ as
having a well-defined free energy, which is however distinct from that of the
corresponding equilibrium state. Nevertheless, it is also clear that
such states eventually decay through a process of nucleation which
brings the metastable state to a true equilibrium state which is
physically quite different.

Traditionally, there have been many attempts to justify associating
some equilibrium state to metastability through analytic continuation
of the free energy. The oldest of these is, of course, the one due to
van der Waals, which indeed works perfectly for mean-field theory. For
models with short-range interactions, however, matters are more
complicated: it has been shown \cite{lan67} that the free energy has
an essential singularity at the coexistence curve. Langer \cite{lan67}
provided a way to define an appropriate analytical continuation
across the existing cut and gave a clear picture of the cause of the
singular behaviour: it is, in fact, due to the presence of the
droplets which eventually nucleate the equilibrium phase.

In what follows we pursue a purely dynamical point of view of
this problem. That is, we start from the main dynamical features of a
metastable state and suggest a reasonable definition in terms of
dynamic features alone. We then show that the metastable state thus
defined can indeed be viewed as the restriction of the equilibrium
ensemble onto a suitably defined subset. To achieve this goal,
however, we must pay a (considerable) price: we must assume the system
obeys very simplified dynamics, namely Markov chains satisfying
detailed balance. Further, we shall  concern ourselves strictly with finite
systems and shall not take the thermodynamic limit. In part this is
due to the fact that real difficulties arise when this limit is taken:
since arbitrarily unlikely fluctuations will arise in arbitrarily
short times in a sufficiently large system, it turns out that
nucleation eventually becomes instantaneous in the thermodynamic
limit, which is clearly an artefact.  A typical way around this
problem might be to take a simultaneous limit to systems of infinite
size as well as to parameters ever closer to the coexistence curve. We
shall instead simply study the finite, though large, system.

The phenomenon of metastability may be described informally as follows
(see \cite{pen71, pen79} for a much fuller discussion along similar
lines): a system is said to be in a metastable state if, upon starting
the system in a certain subset of initial conditions, the system
remains for a very long time in this limited subset of the set of all
configurations, which is of negligible measure in equilibrium.
Further, this subset is macroscopically distinct from the equilibrium
state. Also, the return to equilibrium from a metastable state usually
occurs in an abrupt fashion, i.e.\ the macroscopic variables do not
change slowly from their metastable values to their equilibrium
values, but rather, they remain essentially constant and then suddenly
relax to their equilibrium value, by some quick relaxation
mechanism. The requirement that the time during which the system
remains in a metastable state be ``large'' means simply that it is
sufficient to allow the system to relax to some kind of
pseudo-equilibrium state. Thus, in a metastable state, the values of
the macroscopic observables of interest will not show any systematic
time-dependence, at least after some initial transient, the duration
of which is much less than the decay time of the metastable state.

The results presented in this work are derived for generic ergodic,
acyclic Markov processes satisfying detailed balance with respect to the
Gibbs measure. The typical system we have in mind is the
finite kinetic 2D Ising model, which we discuss in section III.

\section{General formalism}

Let us consider a Markov chain on a finite state space $\Gamma$
with rates $W_{\ssp}$, where $\sigma$ and $\sigmap$ denote states of
$\Gamma$. The master equation for the probability of the system being
found in state $\sigma$ is given by
\begin{eqnarray}
\partial_t P(\sigma)&=&LP;\nonumber\\
LP&=&\sum_\sigmap W_\sps P(\sigmap)-
P(\sigma)\sum_\sigmap W_\ssp.
\label{eq:1.1}
\end{eqnarray}
Since the Markov chain is assumed to be ergodic and acyclic
\cite{bre99}, well-known theorems assert that the solution approaches
a unique equilibrium $P_0(\sigma)$.  If we further assume that
detailed balance holds, that is,
\begin{equation}
W_\sps P_0(\sigmap)=W_\ssp P_0(\sigma),
\label{eq:1.2}
\end{equation}
then it is also well known that the operator $L$ defined in 
(\ref{eq:1.1}) is self-adjoint with respect to the scalar product
\begin{equation}
\left(
\Phi, \Psi
\right)=\sum_\sigma\frac{\Phi(\sigma)\Psi(\sigma)}{P_0(\sigma)}.
\label{eq:1.3}
\end{equation}
Since the underlying vector space is finite-dimensional, it then
follows that there is a complete orthonormal set of eigenvectors $P_n$
satisfying
\begin{equation}
LP_n=-\Omega_nP_n,
\label{eq:1.4}
\end{equation}
where the $\Omega_n$ are by definition arranged in increasing order.
The existence of an equilibrium distribution implies that
$\Omega_0=0$ and the corresponding $P_0$ is in fact the equilibrium
distribution. All other $\Omega_n$ are strictly positive.
                                        
Using the orthonormality of the $P_n$ we can write
\begin{equation}
\sum_\sigma P_n(\sigma)=\delta_{n,0},
\label{eq:1.5}
\end{equation}
implying that $P_0(\sigma)$ is normalized and that adding to it
arbitrary multiples of $P_n(\sigma)$, for $n\geq1$, does not alter
this normalization.
                                        
One then arrives using standard techniques \cite{lar05} at a formal
expression for the probability of arriving from $\sigmaz$ to $\sigma$
in time $t$:
\begin{equation}
P(\sigma,t;\sigmaz,0)=P_0(\sigma)+\sum_{n=1}^\infty
\frac{P_n(\sigma)P_n(\sigmaz)}{P_0(\sigmaz)}
e^{-\Omega_n t}.
\label{eq:1.7}
\end{equation}
We now turn to the characterization of a metastable state within
the general setting outlined above. In view of the informal
description of metastability sketched in the introduction, it is clear
that if we wish to have a behaviour different from equilibrium over a
large time scale, one needs that at least one of the $\Omega_n$ be
much closer to zero than the rest.
                                        
Let us assume that $\Omega_1 \ll\Omega_n$ for all $n\geq2$. Now
consider a process evolving from the initial condition
$\sigmaz$. Then, following (\ref{eq:1.7}), in the relevant time range
$\Omega_2^{-1}\ll t\ll\Omega_1^{-1}$, one finds that the configuration
$\sigma$ is occupied with the following (time-independent) probability
\begin{equation}
P(\sigma)=P_0(\sigma)+\frac{P_1(\sigmaz)}{P_0(\sigmaz)}P_1(\sigma).
\label{eq:1.8}
\end{equation}
Note that, due to (\ref{eq:1.5}), this is normalized. Since it differs
exponentially little from the exact result, we may also conclude that
it is everywhere positive, except perhaps in some places where it
assumes exponentially small negative values; the latter can be fixed
by setting the offending negative values to zero and recomputing the
normalization.

This result focuses our attention on the value
$P_1(\sigmaz)/P_0(\sigmaz)$, which characterizes the nature of the
initial condition. This quantity will be central to understanding the
conditions under which the initial condition can truly be called
metastable and the resulting probability distribution given by
(\ref{eq:1.8}) can justifiably be identified with that of a metastable
state. Let us be more specific.

In what follows, we denote $P_1(\sigma)/P_0(\sigma)$ by $C(\sigma)$,
and the maximum value of $C(\sigma)$ by $C$. Next we
define the two sets $\Gamma_m$ and $\Gamma_{eq}$ as follows:
\begin{equation}
\Gamma_m:=\left\{\sigma:\frac{C}{2}\leq C(\sigma)
\leq C
\right\},
\label{eq:1.9}
\end{equation}
and $\Gamma_{eq}$ is defined to be the complement of $\Gamma_m$. 
The choice of the factor $1/2$ to define the lower bound on $C(\sigma)$ in
(\ref{eq:1.9}) is purely arbitrary and a matter of convention.

We will show that given the previous scenario, the system will have
a metastable state, in the sense discussed in the introduction, if
\begin{equation}
\sum_{\sigma\in\Gamma_m}P_0(\sigma)\ll1,
\label{eq:1.10}
\end{equation}
i.e.\ that the probability of being found in $\Gamma_m$ in equilibrium
is negligibly small, and we define the ``metastable state'' as
the state described by the probability distribution
\begin{equation}
P_m(\sigma)=P_0(\sigma)+CP_1(\sigma).
\label{eq:1.91}
\end{equation}
It should be stressed that, from a physical point of view, condition
(\ref{eq:1.10}) is the {\em crucial\/} assumption: it allows to
distinguish true metastable states from other slowly decaying
states. Of course, in concrete instances this hypothesis will not be
easy to prove rigorously, and, for this reason, our approach is in a
sense somewhat formal. We shall, however, show that a large number of
consequences follow from (\ref{eq:1.10}). It is therefore sufficient
to prove (\ref{eq:1.10}) to show that the restricted state approach to
the statistical description of metastable states is applicable. (Note
that the importance of (\ref{eq:1.10}) was already pointed out  in
\cite{pen71, pen79}.)
                    
In what follows we will show that the properties of systems in which
assumption (\ref{eq:1.10}) holds give rise to a behaviour
which can be identified as metastabilty. These properties are:
                                        
\begin{enumerate}
                                        
\item The probability that a state evolving from an initial condition
$\sigma_0$ for which $C(\sigma_0)=C$ (or very close to it) leaves
$\Gamma_m$ in a time less than $t$ is of order $\Omega_1t$. This
justifies identifying such a state as a very persistent
one. From this result it also follows that
\begin{equation}
\sum_{\sigma\in\Gamma_{eq}}\left[P_0(\sigma)+CP_1(\sigma)\right]\ll1.
\label{eq:1.11}
\end{equation}
From this inequality and the positivity properties discussed above, we
conclude that
\begin{equation}
P_1(\sigma)\approx-C^{-1}P_0(\sigma),\qquad\qquad\sigma\in\Gamma_{eq}.
\label{eq:1.12}
\end{equation}
Note, however, that the above feature is not enough to
characterize a metastable state. A slowly decaying hydrodynamic mode,
say, would have the same property.
\item
The probability that a state is found in $\Gamma_{eq}$ after a time of
order $\Omega_2^{-1}$, evolving from an initial condition $\sigma_0$
such that $C(\sigma_0)=(1-p)C$, is $p$. On the other hand, if the
state has remained in $\Gamma_m$ for a time of order $\Omega_2^{-1}$,
then the value of $C(\sigma_t)$ grows to values very close to $C$ on
the same time scale. These results are crucial, because they mean
that systems which have $C(\sigma_0)\neq C$ relax fast 
either to equilibrium or to the metastable state.
Once they are in the metastable state, they can be described by the
probability distribution $P_m(\sigma)$ defined in (\ref{eq:1.91}).
In order to prove this characteristic property, we  
have to make use of the defining property of metastable states
(\ref{eq:1.10}). 
\item
Finally, if we define a new process in which all transition rates
connecting the metastable region $\Gamma_m$ defined by (\ref{eq:1.9})
to $\Gamma_{eq}$ are set equal to zero, we obtain another Markov
process, also satisfying detailed balance with respect to the
restriction of $P_0(\sigma)$ to $\Gamma_m$. We show that if both
processes are started from the same initial condition $\sigma_0$
satisfying $C(\sigma_0)=C$ then the two processes remain close (in the
sense of distance in variation) over a time of order $\Omega^{-1}$.
This result leads to
\begin{equation}
P_1(\sigma)\approx C P_0(\sigma),\qquad\qquad~~\sigma\in\Gamma_m
\label{eq:1.14}
\end{equation}
and
\begin{equation}
2\ln C=\ln\sum_{\sigma\in\Gamma_{eq}} P_0(\sigma)-
\ln\sum_{\sigma\in\Gamma_m} P_0(\sigma),
\label{eq:1.15}
\end{equation}
which is interpreted in a natural way as the free energy difference
between the two phases.

Note that this final result also allows to carry over standard 
results valid for equilibrium systems to the metastable case: 
one first applies the result to the restricted process, 
which is a {\em bona fide\/} Markov process defined on $\Gamma_m$
for all times, and then extends it to the metastable case by
arguing that the two processes are close for the relevant
timescale $\Omega_2^{-1}\ll t \ll\Omega_1^{-1}$. In particular, the result
derived in \cite{bae03} can partly be rederived in this way: the 
fluctuation--dissipation theorem holds in metastable states of
the kind we describe, because it can be derived as a general
property of Markov processes with detailed balance. 
\end{enumerate}
The above results therefore indicate that the program of defining a
restricted equilibrium ensemble to describe metastability can be
carried out in a fairly rigorous fashion in the context of Markovian
proceses satisfying detailed balance.

For a detailed derivation of these results, see \cite{lar05, lar06}.
Here we content ourselves with a rough sketch of how they come about.
Note first the following basic property:
\begin{equation}
E\left(
\left.e^{\Omega_1t^{\prime}}C(\sigma(t^\prime))\right|\sigma(t)\right)
=e^{\Omega_1t}C(\sigma(t))\qquad(t<t^\prime),
\label{eq:1.16}
\end{equation}
where $\sigma(t)$ denotes a path of the Markov process defined by
(\ref{eq:1.1}), and $E$ denotes the expectation value. This relation
is easily verified by a straightforward computation and means that
$e^{\Omega_1t}C(\sigma(t))$ is a martingale.

To prove point (1), assume that the initial condition $\sigma_{0}$
satisfies $C(\sigma_{0})=C$.  Now (\ref{eq:1.16}) means that, on
average, $C(\sigma) e^{\Omega_1t}$ should neither go up nor
down. Since it starts at the highest possible value of $C(\sigma)$, it
has nowhere to go but down (on short time scales this is not
significantly changed by the factor $e^{\Omega_1t}$). Therefore going
down a significant amount is unlikely. It is therefore not likely to
leave $\Gamma_{m}$ in the relevant time scale.

Point (2) is more technical: it can be shown that the
condition (\ref{eq:1.10}) implies that $P_{0}$ and $P_{0}+CP_{1}$ are
substantially different from zero on two disjoint sets. Therefore, if
the initial condition $\sigma_{0}$ satisfies $C(\sigma_0)=(1-p)C$, it
evolves into a state given by
\begin{eqnarray}
  P^{(p)}(\sigma) & = & P_{0}(\sigma)+(1-p)CP_{1}(\sigma)
  \nonumber \\
   & = & pP_{0}(\sigma)+(1-p)\left[
   P_{0}(\sigma)+CP_{1}(\sigma)
   \right].
  \label{eq:1.17}
\end{eqnarray}
But this state can be interpreted as being in the equilibrium state
with probability $p$ or in the metastable state with probability
$1-p$. 

For point (3) consider the restricted process, where the rates are defined by
\begin{equation}
W^R_\sps=
\cases{W_\sps&\qquad$\sigma,\sigmap\in\Gamma_m$ or
$\sigma,\sigmap\in\Gamma_{eq}$\cr 0&\qquad\mbox{otherwise.}}
\label{eq:1.18}
\end{equation}
This process satisfies detailed balance with respect to $P_{0}$, just
as does the original process. The trajectories of the physical process
starting from $\sigma_{0}$ with $C(\sigma_0)=C$ have the same
probabilities as the corresponding trajectories of the restricted
process \emph{except when the former cross from $\Gamma_{m}$ to
$\Gamma_{eq}$}. But, as was shown in point (1), such crossings are
unlikely, so the two processes indeed remain close to each other in
the relevant time range.

\section{An application: the Ising model}

We now proceed to discuss how these ideas can be applied concretely to
the case of the 2D kinetic Ising model. As is well known, if $T<T_{c}$
and a small magnetic field $h$ is applied, then the spontaneous
magnetization in equilibrium points in the direction of the
field. There is, however, for a broad range of parameters, a
metastable state for which the magnetization is in the direction
opposite to the field.
\begin{figure}
\begin{center}
    \includegraphics{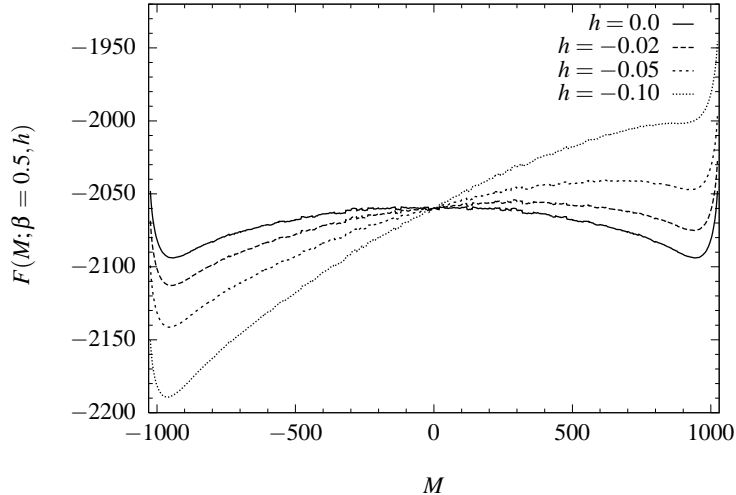}
\end{center}
\caption{ \label{fig:1}
   Free energy $F(M; \beta, h)$ of the 2D Ising model 
    (see (\ref{eq:3.5}) for the definition) for a 
    $32\times32$ sample with $\beta:=1/T=0.5$, i.e.\ significantly below
    criticality. Note the secondary minima for $0\neq|h|<0.05$, 
    which are quite pronounced in spite of our considering a fairly 
    large system.}
\end{figure}

We wish to show a way to obtain some confirmation of the ideas
described above through simulations of this system. A crucial issue is
to identify the observables which play an essential role in
determining $C(\sigma)$ and hence $\Gamma_{m}$. 
For the ferromagnetic Ising model, we assume that these observables
reduce simply to the spin interaction energy $E(\sigma)$ and the magnetization
$M(\sigma)$. For systems of the size we consider presently, it is not possible
for a nucleating droplet, which is the crucial factor determining
whether a system is or is not about to nucleate, to appear without
noticeably affecting the values of $E$ and $M$. We therefore assume that
\begin{equation}
    C(\sigma)=\Phi\left[
    E(\sigma),M(\sigma)
    \right],
    \label{eq:3.1}
\end{equation}
which \emph{defines} $\Phi(E,M)$.

The equilibrium probability of a configuration $\sigma$ is
\begin{equation}
    P_{0}(\sigma)\propto \exp\left[
    -\beta E(\sigma)+\beta hM(\sigma)
    \right],
    \label{eq:3.2}
\end{equation}
where $\beta := 1/T$.  Summing over all configurations $\sigma$ having
the same value of $E$ and $M$, the equilibrium probability that the
system is in the macrostate $(E,M)$ is 
\begin{equation}
    P_{0}(E,M)\propto g(E, M) \exp\left[
    -\beta(E-hM)
    \right].
    \label{eq:3.3}
\end{equation}
Here, $g(E, M)$ is the density of states of the Ising model, given by
\begin{equation}
    g(E, M)=\sum_{\sigma}\delta\left[
    E-E(\sigma)
    \right]
    \delta\left[
    M-M(\sigma)
    \right].
    \label{eq:3.4}
\end{equation}
This can be computed numerically in an efficient manner, say
using the Wang--Landau algorithm \cite{wan01a, wan01b}. 

In the metastable region $\Gamma_{m}$ one has $P_m=P_{0}+C
P_{1}=(1+C^2)P_{0}$, so that $P_m$ can also be expressed in terms of
$E$ and $M$ using (\ref{eq:3.3}).  Since the metastable state is well
characterized by specific values of $E$ and $M$, it is to be expected
that the expression on the r.h.s.\ of (\ref{eq:3.3}) will show a local
maximum at these non-equilibrium values.  This is confirmed
numerically for $P_0(E,M)$ as a function of $E$ and $M$ \cite{lar06}.

Here we plot in Figure \ref{fig:1} the free energy
\begin{equation}
    F(M; \beta,h):=-\frac{1}{\beta}\ln\sum_{E}g(E,M) \exp[-\beta(E-hM)]
    \label{eq:3.5}
\end{equation}
at a fixed subcritical temperature.  A secondary minimum is seen in
Figure \ref{fig:1}, which corresponds to the metastable maximum of
$P_0$. 
Such a state of affairs will not exist in the thermodynamic
limit: indeed, up to an additive term $hM$, $F(M; \beta,h)$ is the
free energy of the Ising model in an ensemble of fixed temperature and
magnetization. By standard theorems on short-range systems
\cite{rue99}, this must be equivalent in the thermodynamic limit to
the free energy computed as the Legendre transform of the Gibbs
potential calculated in the grand canonical ensemble.  Hence, as the
thermodynamic limit is approached, $F(M; \beta,h)$ must tend to a
convex function in $M$, and the additive term $hM$ does not alter this
fact. This is clearly at odds with the presence of two minima in
$F(M; \beta,h)$ as shown in Figure \ref{fig:1}. Therefore, our
identification of $E$ and $M$ as adequate variables to determine
$C(\sigma)$ is not tenable beyond a certain sample size. This is in
agreement with the fact that for large systems a nucleating droplet
may appear without affecting the values of $E$ and $M$.

In the systems in which $E$ and $M$ do furnish a complete description,
it is possible to determine $C$ and hence the free
energy difference betwen the stable and metastable phase as follows:
(\ref{eq:1.15}) yields $C$ as a function of the probability of
finding the system in $\Gamma_{m}$ in equilibrium. Since we have
identified $\Gamma_{m}$ with a certain part of $(E,M)$ space, we can
readily compute this probability once $g(E,M)$ is known. This free
energy difference can also be compared with one obtained from
hysteresis curves, and the comparison is quite satisfactory. These results
will be discussed more extensively in \cite{lar06}.
Finally, extensions of this formalism to systems characterized by
having several metastable states appear possible and are also
presently under way.

\begin{acknowledgments}
  The support of DGAPA project IN100803 is gratefully acknowledged. 
  Incisive comments from P.~H\"anggi, which contributed to the
  authors' understanding of the role of the thermodynamic limit, 
  are also sincerely appreciated. 
\end{acknowledgments}

\end{document}